\documentclass[12pt]{article}
\usepackage{graphics}
\usepackage{amssymb}
\usepackage{color}
\usepackage{amsmath}
\usepackage{amsthm}
\usepackage{amsopn}
\usepackage{axodraw}
\def\uno{\mbox{1 \kern-.59em {\rm l}}}

\def\p{\partial}

\def\th{\theta}
\def\nn{\nonumber}
\def\be{\begin{equation}}
\def\ee{\end{equation}}
\def\bea{\begin{eqnarray}}
\def\eea{\end{eqnarray}}

\begin{document}

 \begin{center}

 {\bf{\large Two-photon annihilation of singlet cold dark matters due to noncommutative space-time}}

\vskip 4em

 {{\bf M. M. Ettefaghi} \footnote{ mettefaghi@qom.ac.ir  }% and {\bf H. Haj-Mola-Heydar}
 }
 \vskip 1em
 Department of Physics, The University of Qom, Qom 371614-6611,
Iran.

 \end{center}

 \vspace*{1.9cm}

\begin{abstract}
Detecting the cosmic rays, in particular gamma-ray, coming from the
dark matter annihilation or decay is an indirect way to survey the
nature of the dark matter. In the commutative space-time, the
annihilation of the dark matter candidates (WIMPs) to photons
proceeds through loop corrections. However, it is possible for WIMPs as well as the other
standard model singlet particles to couple with photons
directly in the noncommutative space-time. In this paper, we study
two-photon annihilation of singlet WIMPs in the noncommutative space-time. If the noncommutative interactions are relevant to the relic abundance, one can exclude some dark matter masses using Fermi-Lat data.
\end{abstract}
\newpage
%%%%%%%%%%%%%%%%%%%%%%%%%%%%%%%%%%%%%%%%%%%%%%%%%%%%%%%%%%%%%%%%%%%%%%%%%%%%%%%%%%%%%%%%%%%%%%%%%%%%%%%%%%%%%%%%%%%%%%%%%%%
%\newpage
\section{Introduction}
Detection of the annihilation of dark matters into monochromatic
gamma rays in upcoming telescopes is certainly an appropriate way to
unambiguously determine their unknown nature. The most papular
candidates of dark matter are the weakly interacting massive
particles (WIMP) which are accommodated in some models beyond the
standard model (SM) such as supersymmetry models with R parity
\cite{rev1,rev2}, the extra dimensional models with conserved
Kaluza-Klein (KK) parity \cite{kk}, the T-parity conserved little
Higgs model \cite{little higgs}, and so on. Also singlet particles, either
scalars \cite{scalar} or fermions \cite{fermion1,fermion2,q},
can be served as cold dark matter. In all above scenarios,
the weak interactions of WIMPs are the main key to explain the
thermal production of them in the early universe ( for review see
\cite{rev1,rev2}). Additionally, these weak interactions can provide
an opportunity to search dark matters through their production in
high energy accelerators \cite{peskin}, their direct detection
\cite{direct}, and their indirect detection, i.e. astrophysical
observations of their annihilation or decay products in our galaxy
or beyond. In fact, through the WIMP scenario, the weak interaction
of dark matters would produce observable SM particles,
such as charged anti-matter particles, photons  and neutrinos. Among
these, neutrino and photons have advantage in comparison to others,
because they keep their source information during the streaming.
Moreover, the very small cross sections of neutrinos make their flux
very difficult to detect. Therefore, the gamma ray signatures of the
dark matter have been investigated extensively (For review see
\cite{gamma rev} and references therein). The continuum gamma ray
emission from dark matter annihilation could be confused with
astrophysical backgrounds, e.g. emission from galactic cosmic rays or
from milli-second pulsars. Hence, the study of the monochromatic gamma ray
is important. Monochromatic gamma ray signatures have been studied
for some dark matter candidates in literature \cite{scalargammaray,gammaray}.

On the other hand, noncommutative (NC) quantum field theories have
been considered in the recent decade extensively because of some
motivations coming from string theory \cite{douglas} and measurement arguments based on quantum mechanics and classical gravity \cite{minimallength}.  In the NC field theory, one encounters
new properties such as UV/IR mixing problem \cite{UVIR}, the violation of Lorentz invariance.\footnote{One can see from (\ref{ncalgebra}), NC parameter, $\th^{\mu\nu}$, is a constant antisymmetric matrix which specifies a prefer direction in the space-time. However, quantum field theory on NC space-time possesses
symmetry under a twisted Poincar\'{e} algebra whose representation content is identical to the usual Poincar\'{e} symmetry \cite{lorentz}.} %and CP-violation \cite{charge}.
 From the phenomenological point of view, by
comparing the results of noncommutative version of usual
physical models with present data, lower bounds on the
noncommutative scale have been estimated conservatively
about 1-10 TeV \cite{bound}.
The NC field theories are constructed on the space-time coordinates,
which are operators and do not obey commutative algebra. In the case
of canonical version of the NC space-time, the coordinates satisfy
the following algebra: \be \label{ncalgebra}[\hat{x}^\mu,\hat{x}^\nu]=i\th^{\mu\nu},
\ee where a hat indicates a NC coordinate and $\th^{\mu\nu}$ is a
real, constant and antisymmetric matrix.
  According to the Weyl-Moyal correspondence, to construct the NC field theory, an ordinary function can
  be used instead of the corresponding NC one by replacing the ordinary product with the star product as follows:
\be\label{starproduct}
 f\star
g(x,\theta)=f(x,\theta)\exp(\frac{i}{2}\overleftarrow{\partial}_\mu
\theta^{\mu\nu}\overrightarrow {\partial}_\nu)g(x,\theta). \ee Due
to the above correspondence, a neutral particle (as well as a charged
particle) can couple with the $U(1)$ gauge field in the adjoint
representation. Some effects of this new
coupling were studied in the literature \cite{nph}. In particular, singlet particles, which can be
served as cold dark matter, can couple with the $U(1)$ electroweak
gauge field in this manner \cite{q}. For instance, this interaction can be relevant to the production of dark matters with masses about 100 GeV provided that the NC scale is about the 1 TeV.

In the usual space time, there does not exist any model which predicts direct coupling between WIMPs and photons, because they are electrically neutral. Hence the annihilation of WIMPs into photons proceeds through loop corrections. However, this process is possible at the tree level for the standard model singlet particles through adjoint representation of $U(1)$ gauge theory in the NC space-time. In this paper, we calculate the annihilation cross section
of singlet dark matters into two photons in the NC space-time. Although this proceeds at the tree level, its contribution is suppressed with $\th^4$ or
equivalently $\frac{1}{\Lambda^8}$ where $\Lambda$ is the NC scale. However, this NC induced interaction can be relevant to the thermal production of singlet dark matters in some parameter regions \cite{q}. Therefore, the study of this process helps one to constraint the corresponding parameter regions using  gamma-ray experiments such as Fermi-Lat \cite{fermi}.

This
paper is organized as follows: In Sec. II we give a brief review of singlet extended noncommutative standard model. In Sec. III and Sec. IV we study the annihilation of singlet fermion and scalar, respectively, into two photons. Finally, we
discuss on our conclusions in Sec. V.

\section{A brief review of singlet extended noncommutative standard model}
The Weyl-Moyal correspondence, Eq. (2), leads to a few restrictions on a gauge theory in the NC space-time \cite{nogo}:
%\begin{itemize}
 (a) Only $U(n)$ gauge theories have a NC extension without any enlargements. Because of existing some terms proportional to the identity matrix due to the Weyl-Moyal correspondence,  usual $SU(n)$ gauge theories in particular are not permissible.
(b)  Only $n\times n$ matrix representations of $u(n)$ algebra respect the closeness condition. For instance, in the
$U_\star(1)$ case for arbitrary fixed charge $q$, only the
matter fields with charges $\pm q$ and zero are permissible.
(c) In a gauge theory consisting of several simple gauge groups, the
matter fields cannot carry more than two NC gauge group charges.
%\end{itemize}
Hence, the extension of the standard model based on $SU(3)\times SU(2)\times U(1)$ gauge theory to the NC space-time is problematic.  There exist, however, two approaches
to construct the standard model gauge theory in the NC
space-time.
%\begin{itemize}
%\item

In the first approach, the gauge group is restricted
to $U(n)$ and the symmetry group of the standard model is
achieved by the reduction of $U(3)\times U(2)\times U(1)$ to
$SU(3)\times SU(2)\times U(1)$ by an appropriate symmetry
breaking \cite{nNCSM}. Namely, two extra $U(1)$ factors are reduced through two extra Higgs particles (rather than the standard model) during appropriate Higgs mechanisms. The number of possible particles in each family is six; left-handed leptons, right-handed charged leptons, left-handed quarks, right-handed up quarks, right-handed down quarks, and Higgs which transform under the standard model gauge group as follows:
\bea
\Psi^l_L(x)&&\hspace{-5mm}\equiv\Big(\begin{array}{c}
                   \nu(x) \\
                   e(x)
                 \end{array}
\Big)_L\rightarrow V(x)\star\Psi^l_L(x)\star v^{-1}(x)\\
e_R(x)&&\hspace{-5mm}\rightarrow e_R(x)\star v^{-1}(x)\\
\Psi^q_L(x)&&\hspace{-5mm}\equiv\Big(\begin{array}{c}
                   u(x) \\
                   d(x)
                 \end{array}
\Big)_L\rightarrow V(x)\star\Psi^q_L(x)\star U^{-1}(x)\\
u_R(x)&&\hspace{-5mm}\rightarrow v(x)\star u_R(x)\star U^{-1}(x)\\
d_R(x)&&\hspace{-5mm}\rightarrow d_R(x)\star U^{-1}(x)\\
H(x)&&\hspace{-5mm}\equiv\Big(\begin{array}{c}
                   H^+(x) \\
                   H^0(x)
                 \end{array}
\Big)\rightarrow V(x)\star H(x),
\eea
where $v(x)$, $V(x)$, and $U(x)$ are $U(1)$, $U(2)$ and $U(3)$ gauge transformations, respectively.
These transformations along with the following transformation for the gauge fields:
\bea
B_\mu\rightarrow v\star B_\mu\star v +\frac i g_1 v\star\p_\mu\star v, \\
W_\mu\rightarrow U\star W_\mu\star U +\frac i g_2 U\star\p_\mu\star U, \\
G_\mu\rightarrow V\star G_\mu\star V +\frac i g_3 V\star\p_\mu\star V,
\eea
where $B_\mu$, $W_\mu$ and $G_\mu$ are $U(1)$, $U(2)$ and $U(3)$ gauge fields, respectively, define the $U(3)\times U(2)\times U(1)$ gauge theory including gauge and Yukawa interactions. Of course, the conservation of the gauge symmetry in the Yukawa interactions for the up quarks leads to the following gauge transformation for the charge conjugated of doublet Higgs field:
\be
H^C\rightarrow V(x)\star H^C\star v^{-1}.
\ee As was said, in addition to the standard model contents, there exist two new Higgs to reduce two addition $U(1)$ factors. The details of this model is out of the scope of the recent paper (for the detail of model building see \cite{nNCSM}).  Moreover, a singlet particle either fermion or scaler with the following gauge transformation:
\be\label{singlet}
\Phi(x)\rightarrow v(x)\star\Phi(x)\star v^{-1}(x),\ee
can be accommodated. The gauge interactions of these particles as well as Yukawa coupling between singlet fermions and scalars are permissible. However, the couplings between standard model Higgs and singlet scalar, which is possible in the usual space-time, violate the gauge symmetry in this model \cite{q}.
%\item

In the second approach, one can construct the $SU(n)$ gauge group in the noncommutative space-time using Seiberg-Witten maps \cite{swmap}.
In this manner, a version of the NC standard model has been constructed which includes only the content of the usual one \cite{sm}\footnote{We should mention that  in the  works \cite{swmap} and \cite{sm}, the charge quantization problem  inherent in NC gauge field theories  discussed and  treated in \cite{nogo} and \cite{nNCSM}, was dismissed  by "mapping" three different noncommutative gauge field degrees of freedom to a single ordinary gauge field \cite{bypass}.}. Explicitly, the Lagrangian of this theory is similar to the commutative standard model, but the fields and products are replaced by the NC fields and star products, respectively. For the practical purposes, the NC fields have to be written with respect to the usual fields using Seiberg-Witten maps. Although the interactions of the standard model receive the NC corrections, one encounters some new interactions between the gauge fields themselves or between gauge fields and matter fields which proceed through loop corrections in the commutative space-time \cite{sm}.
In this approach,
a singlet particle, either fermion or scalar, can also be transformed under gauge transformation according to (\ref{singlet}). Therefore, it can be involved in the gauge interactions through the following minimal coupling:
 \be
\hat{D}_\mu\hat{\Phi}=\p_\mu\hat{\Phi}-ig^\prime(\hat{B}_\mu\star\hat{\Phi}-\hat{\Phi}\star
\hat{B}_\mu),\ee where hats on the fields are used to emphasize that
these fields are defined in the NC space-time. It is clear that the Yukawa coupling
between the singlet fermion and the singlet scalar
is gauge invariant. In addition, the interaction terms
between S and the standard model Higgs doublet H, such
as $H^\dagger HS$ and $H^\dagger HS^2$, do not violate the gauge symmetry
if H transforms under the following representation \cite{q}:
\be H\rightarrow V\star H\star v^{-1}.\ee

Therefore, a singlet particles beyond the standard model can be coupled with the $U(1)$ gauge field in both versions of the NC standard model.
Using Seiberg-Witten maps and Weyl-Moyal correspondence, we expand the relevant action to the annihilation of singlet particles into two photons in terms of the NC parameter, $\th$. To obtain the lowest order of NC corrections, we need the following Seiberg-Witten maps of the singlet particles either fermion or scalar
\cite{swmap}: \bea\label{sw1}
\hat{\Phi}=&&\!\!\!\!\!\!\!\Phi+g^\prime\th^{\mu\nu}B_\nu\p_\mu\Phi+{g^\prime}^2\th^{\mu\nu}\th^{\kappa\lambda}[\frac 1 2 B_\mu B_\kappa\p_\nu\p_\lambda\Phi
-\p_\mu B_\kappa B_\nu\p_\lambda\Phi+\nn\\
&&\frac 1 4 B_\nu\p_\kappa B_\mu\p_\lambda\Phi+\frac 1 8 \p_\mu B_\kappa\p_\nu B_\lambda\Phi+\frac 1 8 \p_\kappa
B_\nu\p_\mu B_\lambda\Phi],\eea
and of the $U(1)$ gauge field:
\be\label{sw2}
\hat{B}_\mu=B_\mu+e\th^{\nu\rho}B_\rho[\p_\nu
B_\mu-\frac 1 2 \p_\mu B_\rho].\ee

\section{Annihilation of singlet fermions into two photons}
In the commutative space-time, the singlet particles interact with the standard model particles through a renormalizable coupling between singlet scalar and the standard model Higgs. As was said, the transcription of this coupling in the NC space-time leads to the violation of $U(3)\times U(2)\times U(1)$ gauge symmetry. Therefore, the singlet particle production is only possible to be explained by using the NC induced interaction in the $U(3)\times U(2)\times U(1)$ model. Moreover, if one can consider the coupling between the singlet scalar and the standard model Higgs, the NC contribution in the annihilation of singlet particles to photons is comparable or larger than the commutative one for some region of parameter space. That is why it proceeds at the tree level in the NC space-time (Fig. \ref{fig1}) in contrary to the commutative one which proceeds through loop quantum corrections.
\begin{figure}[h]
\vspace*{-5mm}
\[
  \vcenter{\hbox{
  \begin{picture}(130,80)(-10,0)
  \ArrowLine(5,15)(30,40)
   \ArrowLine(60,40)(85,15)
  \Photon(30,40)(5,65){2}{5}
   \Photon(60,40)(85,65){2}{5}
 % \ArrowArc(65,50)(25,-180,0)
  \ArrowLine(30,40)(60,40)%{2}
  %\Line(60,40)(75,55)
   \Text(10,27)[b]{\scriptsize{$\Phi$}}
  \Text(80,27)[b]{\scriptsize{$\bar\Phi$}}
  \Text(21,57)[b]{\scriptsize{$\gamma$}}
  \Text(70,57)[b]{\scriptsize{$\gamma$}}
    \end{picture}
  \begin{picture}(230,80)(-30,0)
  \ArrowLine(5,15)(30,40)
  \Photon(30,40)(60,65){2}{5}
   \Photon(65,40)(35,65){2}{5}
 % \ArrowArc(65,50)(25,-180,0)
  \ArrowLine(30,40)(65,40)%{2}
    \ArrowLine(65,40)(90,15)
  \Text(10,27)[b]{\scriptsize{$\Phi$}}
  \Text(83,27)[b]{\scriptsize{$\bar\Phi$}}
          \Text(65,60)[b]{\scriptsize{$\gamma$}}
  \Text(30,60)[b]{\scriptsize{$\gamma$}}
  \end{picture}
   \begin{picture}(330,80)(60,0)
  \ArrowLine(5,15)(30,40)
  \Photon(30,40)(55,65){2}{5}
   \Photon(30,40)(5,65){2}{5}
 % \ArrowArc(65,50)(25,-180,0)
  \ArrowLine(30,40)(55,15)%{2}
     \Text(10,27)[b]{\scriptsize{$\Phi$}}
  \Text(49,27)[b]{\scriptsize{$\bar\Phi$}}
          \Text(43,60)[b]{\scriptsize{$\gamma$}}
  \Text(17,60)[b]{\scriptsize{$\gamma$}}
  \end{picture}}}
\]
\vspace*{-10mm}
%\caption[]{The tree level Feynman diagrams for the annihilation of two singlet particles into monochromatic gamma-ray lines in NC space-time. Here, $\Phi$ stands for a singlet either fermion or scalar.}
\label{fig1}
\end{figure}
%\vskip 5pt
%\begin{figure}
%\centerline{\epsfysize=2.5in\epsfxsize=3in\epsffile{fig.eps}}\caption{The tree level Feynman diagrams for the annihilation of two singlet particles into %monochromatic gamma-ray lines in NC space-time. Here, $\Phi$ stands for a singlet either fermion or scalar.
% }
%\label{fig1}
%\end{figure}
 In particular, we are interested in the parameter region where the NC contribution is larger than the commutative one. Hence, we ignore the interferences of commutative and NC terms.

The action describing a
singlet fermion field in the NC space-time is \be S=\int
d^4x(\bar{\hat{\psi}}\star
i\gamma^\mu\hat{D}_\mu\hat{\psi}-m\bar{\hat{\psi}}\star\hat{\psi}).\ee
After using the mentioned Seiberg-Witten maps and expanding the star product
up to the second order of $\th$, we can write the action as
follows: \bea S=\int
d^4x&&\!\!\!\!\!\!\!\!\bar{\psi}[(i\gamma^\mu\p_\mu-m)-\frac e 2
\th^{\nu\rho}(i\gamma^\mu(B_{\nu\rho}\p_\mu+B_{\mu\nu}\p_\rho+B_{\rho\mu}\p_\nu)-mB_{\nu\rho})]\psi\nn\\
&&\!\!\!\!\!\!\!\!+i{g^\prime}^2\th^{\alpha\beta}\theta^{\kappa\lambda}[\p_{\kappa}\bar{\psi}i\p_\alpha B\!\!\!/\p_\beta\psi B_\lambda
-\p_{\beta}\bar{\psi}i\p_\alpha B\!\!\!/\p_\kappa\psi B_\lambda-\p_{\alpha}\bar{\psi}i\p_\kappa B\!\!\!/\p_\beta\psi B_\lambda\nn\\
&&\!\!\!\!\!\!\!\!+\frac 1 2 \p_{\alpha}\bar{\psi}i\p \!\!\!/B_\kappa\p_\beta\psi B_\lambda]
,\eea
where $B_{\mu\nu}=\p_\mu B_\nu-\p_\nu B_\mu$. The second order terms with respect to $\th$ are relevant to the third diagram of Fig. \ref{fig1} which includes only on-shell external lines. Hence, we have not written those terms which are in the order of $\th^2$ and vanish for the on-shell particles. After electroweak symmetry breaking, $B_\mu$ is written in terms of photon and $Z_0$ as follows:
\be B_\mu=-\sin\th_W {Z_0}_\mu+\cos\th_W A_\mu,\ee
in which $\th_W$ is the electroweak mixing angle.
Therefore, the relevant vertices to the fermionic singlet particles annihilation into two photons are given by \be\label{rule1}
\Gamma^\mu(\psi(k^\prime)\bar{\psi(k)}\gamma(q))=-e[q\th k\gamma^\mu+(k\!\!\!/-m_\nu)\tilde{q}^\mu-q\!\!\!/\tilde{k}^\mu],\ee
and
\bea
T^{\mu\nu}=ie^2[&&\hspace{-5mm}k_2\th p_2{\tilde{p}_1}^\mu\gamma^\nu+p_2\th p_1{\tilde{k}_2}^\mu\gamma^\nu-k_2\th p_1{\tilde{p}_2}^\mu\gamma^\nu\nn\\
&&\hspace{-5mm}-k_1\th p_1{\tilde{p}_2}^\nu\gamma^\mu+k_1\th p_2{\tilde{p}_1}^\nu\gamma^\mu+p_2\th p_1{\tilde{k}_1}^\nu\gamma^\mu\nn\\
&&\hspace{-5mm}+\frac 1 2 p_2\th p_1({k\!\!\!/}_1-{k\!\!\!/}_2)\th^{\mu\nu}],
\eea
where we have used $e=g^\prime \cos\th_W$ and the notation $q\th k=\th^{\mu\nu}q_\mu k_\nu$ and $\tilde{q}^\mu=\th^{\mu\nu}q_\nu$.

The cross section of the self annihilation of the singlet fermion into two photons is obtained as follows:
\bea
\sigma v=&&\hspace{-5mm}0.15\times 10^{4}\times \nn\\
&&\hspace{-5mm}\frac{s\alpha^2(2.89sm^6-7.07m^8-0.340s^2m^4+0.067m^2s^3+0.075s^4)}{m^4{\Lambda^8}_{NC}}
\eea
where $s$ and $v$ are the square center of mass energy and relativistic velocity, respectively. The thermal average cross section times the relativistic velocity can be expanded in terms of $\epsilon=\frac{s-4m^2}{4m^2}$ in the case of non-relativistic singlet fermions. The first non-zero term of this expansion corresponds to $\epsilon=0$. Therefore, we have
\be\label{phan}
\langle\sigma v\rangle=1.3\times10^5\alpha^2\frac{m^6}{\Lambda^8_{NC}}.
\ee
One can look at this result from two points of view. First,
since the annihilation of WIMPs into photons proceeds through loop corrections in the usual space-time, the cross section of the annihilation into photons is four or five orders of magnitude weaker than the annihilation cross section requested by the correct relic abundance.
 Therefore, the comparison of the NC result with the usual expectation in commutative space leads to $\Lambda_{NC}$ to be more than 1 TeV. Second, the NC induced singlet fermion-photon interaction may be relevant to the thermal production of singlet fermion in the early universe provided that \cite{q}
\be\label{relic}
\langle\sigma_{ann}v\rangle_{f} \sim 3.9\times 10^{-3}N_f
\frac{m^2}{\Lambda_{NC}^4},
\ee
where $N_f$ denotes the number of
allowed pair charged fermions. Here, $\langle\sigma_{ann}v\rangle_{f}$ is the thermal average of the annihilation cross section of the singlet fermionic dark matters into the standard model massive particles. The correct relic abundance requires $\langle\sigma_{ann}v\rangle_{f} \sim 1.4 \times 10^{-26} cm^3 s^{-1} \simeq 1.2 \times 10^{-9} GeV^{-2}$. Combination of (\ref{phan}) and (\ref{relic}) leads to
\be
\langle\sigma v\rangle \sim 1.9\times 10^{-14}m^2.
\ee
Obviously, $\langle\sigma v\rangle$ for dark matter with mass about 100 GeV is about Fermi-Lat bounds. Consequently, the singlet fermionic dark matter in the NC space-time may be excluded by Fermi-Lat \cite{fermi} for masses larger than 100 GeV.
\section{Annihilation of singlet scalar into two photons}
Alternatively, a real singlet scalar can also be served as cold dark matter \cite{scalar}. The annihilation of this candidate of dark matter into photon has been studied in the usual commutative theory \cite{scalargammaray}. Therefore, it is also interesting to study the gamma ray coming from the annihilation of the recent candidate of dark matter in the NC space-time. The action describing the singlet scalar particles in the NC space-time is written as follows:
\be
S=\frac 1 2\int d^4x((\hat{D}_\mu\hat{\phi})^\dagger\star \hat{D}^\mu\hat{\phi}-m^2\hat{\phi}^\dagger\star \hat{\phi}).
\ee
After replacing the corresponding Seiberg-Witten maps from (\ref{sw1}) and (\ref{sw2}), and the star products up to the first order of $\th$, we have
\bea
S=\int d^4x [&&\hspace{-5mm}\frac 1 2 \p_\mu\phi \p^\mu\phi-\frac 1 2 m^2\phi\phi
+g^\prime\th^{\alpha\beta}\{\p_\beta\phi\p^\mu\phi\p_\alpha B_\mu-\p_\alpha\phi(\Box-m^2)\phi B_\beta\}\nn\\
&&\hspace{-5mm}-2{g^\prime}^2\th^{\alpha\beta}\th^{\kappa\lambda}[\p_\alpha\phi\p_\mu\p_\kappa\phi B_\beta\p^\mu B_\lambda+\p_\alpha\phi\p_\mu\p_\lambda\phi B_\beta\p_\kappa B^\mu\nn\\
&&\hspace{-5mm}+\frac 1 2 \p_\beta\phi\p_\lambda\phi\p_\alpha B_\mu\p_\kappa B^\mu+\p_\alpha\p_\mu\phi\p_\beta\phi B_\lambda(\p_\kappa B^\mu-\frac 1 2 \p^\mu B_\kappa)\nn\\&&\hspace{-5mm} +\p_\beta\p_\mu\phi\p_\kappa\phi\p_\alpha B^\mu B_\lambda],
\eea
 for the real scalar field. Therefore, the vertex of the coupling between the singlet scalar and photon at the first order of $\th$ is given by
\bea
\Gamma^\mu(\phi(k_1)\phi(k_2)\gamma(q))=-e\th^{\alpha\beta}[&&\hspace{-6mm}{k_1}_\beta{k_2}_\alpha{k_2}^\mu+{k_2}_\beta{k_1}_\alpha{k_1}^\mu\nn\\&&\hspace{-6mm}
-{k_1}_\alpha g_\beta^\mu(k_2^2-m^2)-{k_2}_\alpha g_\beta^\mu(k_1^2-m^2)],
\eea
where we use $e=g^\prime\cos\th_W$. At the second order of $\th$ we have
\bea
T^{\mu\nu}=2ie^2[&&\hspace{-7mm}2k_1\th p_1\tilde{p}_2^\nu p_1^\mu+2k_2\th p_2\tilde{p}_1^\mu p_2^\nu+2k_1\th p_2\tilde{p}_1^\nu p_2^\mu+2k_2\th p_1\tilde{p}_2^\mu p_1^\nu\nn\\
&&\hspace{-7mm}-k_1\th p_1k_2\th p_2g^{\mu\nu}-k_2\th p_1k_1\th p_2g^{\mu\nu}-\tilde{p}_1^\mu\tilde{p}_2^\nu(p_1.k_1+p_2.k_2)\nn\\
&&\hspace{-7mm}-\tilde{p}_1^\nu\tilde{p}_2^\mu(p_2.k_1+p_1.k_2)+p_1\th p_2\tilde{k}_2^\mu(p_1^\nu-p_2^\nu)+p_1\th p_2\tilde{k}_1^\nu(p_1^\mu-p_2^\mu)\nn\\&&\hspace{-7mm}+\frac 1 2 p_1\th p_2(k_1.p_1-k_2.p_1)\th^{\mu\nu}+\frac 1 2 p_1\th p_2(k_2.p_2-k_1.p_2)\th^{\mu\nu}],
\eea
where the scalars are considered to be on the mass-shell. Therefore, after a straightforward calculation, we obtain the cross section of the annihilation of singlet scalars into two photons as follows:
\be
\sigma v=2\alpha^2\frac{71m^4s^3-159m^6s^2+150m^8s+291m^{10}-7.49m^2s^4+2.22s^5}{m^4\Lambda_{NC}^8}.
\ee
Similar to the fermionic case, we consider non-relativistic singlet scalars and expand the thermal average of the annihilation cross section times velocity in terms of $\epsilon=\frac{s-4m^2}{4m^2}$. Putting $s=0$, we obtain the first non-zero in the non-relativistic limit as follows
\be\label{sann}
\langle\sigma v\rangle=6.5\times10^3\alpha^2\frac{m^6}{\Lambda_{NC}^8}.
\ee
We also look at this result from two points of view. First, taking $m=100 GeV$ and $\Lambda_{NC}=1 TeV$ leads to $\langle\sigma v\rangle\sim3.5\times10^{-13}(GeV)^{-2}$ which is comparable with the expectations for the corresponding value in non-resonance region in the commutative space-time \cite{scalargammaray}. Second, the NC induced interactions can be relevant to the thermal explanation of relic abundance provided that
\be\label{relic2}
\langle\sigma_{ann}v\rangle_{s} \sim 1.9\times 10^{-2}N_f
\frac{m^2}{\Lambda_{NC}^4},
\ee
where $N_f$ denotes the number of
allowed pair charged fermions. Here, $\langle\sigma_{ann}v\rangle_{s}$ is the thermal average of the annihilation cross section of singlet scalar dark matters into the standard model massive particles. After replacing the value of $\langle\sigma_{ann}v\rangle_{s}$ and combining with (\ref{sann}), we obtain
\be
\langle\sigma v\rangle\sim3.7\times10^{-17}m^2.
\ee
Consequently, in this case, $\langle\sigma v\rangle$ is comparable or larger than Fermi-Lat bounds \cite{fermi} for dark matters with mass larger than about 3 TeV.
\section{Summary and Discussion}
In this paper, we have considered an extension of the standard model which includes a singlet either fermion or scalar as cold dark matter in the NC space-time. The NC space-time induces a new coupling between singlet particles with photon through the adjoint representation. This coupling may be relevant to the thermal explanation of the dark matter production at early universe provided that the NC scale is about 1 TeV which is consistent with the phenomenologically obtained bounds \cite{q}. In this paper, we have calculated the cross section of the annihilation of singlet, either fermion or scalar, into two photons in NC space-time.
We have found if the NC scale is such that the NC induced interactions are relevant to the production of dark matters, the recent Fermi-Lat results \cite{fermi} will exclude the masses larger than 100 GeV for fermionic dark matter and 3 TeV for scalar dark matter.

{\bf Acknowledgement:}
The author would like to thank Mohammad Reza Piroz for reading the manuscript and H. Hajmollaheydar for her contribution to this work
at preliminary stage. The financial support of the University of Qom research council is also acknowledged.


\begin{thebibliography}{99}
\bibitem{rev1}
G. Jungman, M. Kamionkowski, and K. Griets, Phys. Rep. {\bf 267},
195 (1996).
\bibitem{rev2}
G. Bertone, D. Hooper, and J. Silk, Phys. Rep. {\bf 405}, 279
(2005).
\bibitem{kk}
H.-C. Cheng, J. L. Feng, and K. T. Matchev, Phys. Rev. Lett. {\bf
89}, 211301 (2002); G. Servant and T. Tait, Nucl. Phys. {\bf B 650},
391 (2003).
\bibitem{little higgs}
H.-C. Cheng and I. Low, J. High Energy Phys. 09 (2003) 051; J. High
Energy Phys. {\bf 08},  061 (2004).
\bibitem{scalar}
J. McDonald, Phys. Rev. {\bf D 50}, 3637 (1994); Phys. Rev. Lett.
{\bf 88}, 091304 (2002); M. C. Bento, O. Bertolami, and R.
Rosenfeld, Phys. Lett. {\bf B 518}, 276 (2001); H. Davoudiasl, R.
Kitano, T. Li, and H. Murayama, Phys. Lett. {\bf B 609}, 117 (2005);
C. P. Burgess, M. Pospelov, and T. ter Veldhuis, Nucl. Phys. {\bf B
619}, 709 (2001); Wan-Lei Guo, Yue-Liang Wu, JHEP {\bf 1010}, 083 (2010); etc...
\bibitem{fermion1}
K.Y. Lee, Y. G. Kim, and S. Shin, J. High Energy Phys. {\bf 05}, 100
(2008).
\bibitem{fermion2}
Y. G. Kim and K.Y. Lee, Phys. Rev. {\bf D 75}, 115012 (2007).
\bibitem{q}
M. M. Ettefaghi, Phys. Rev. D{\bf 79}, 065022 (2009).
\bibitem{peskin}
E. A. Baltz, M. Battaglia, M. E. Peskin, and T. Wizansky, Phys. Rev.
{\bf D 74}, 103521 (2006).
\bibitem{direct}
M. W. Goodman, and E. Witten, Phys. Rev. D{\bf 31}, 3059 (1985); S.
Khali and C. Munoz, Contemporary Physics {\bf 43}, 51 (2002); C.
Munoz, Int. J. Mod. Phys. A {\bf 19} 3093 (2004).
\bibitem{gamma rev}
Gianfranco Bertone, [arxiv:astro-ph/0607706].
\bibitem{scalargammaray}
C. E.
Yaguna, JCAP {\bf 0903}, 003 (2009) [arxiv:astro-ph/0810.4267]; S. Profumo, L. Ubaldi, and C. Wainwright, Phys. Rev. D{\bf 82}, 123514 (2010) [arxiv:hep-ph/1009.5377].
\bibitem{gammaray}
M. Gustafsson, E. Lundstrom, L. Bergstrom, and J. Edsjo, Phys. Rev.
Lett. {\bf 99}, 041301 (2007) [arxiv:astro-ph/0703512];  F. Takayama and M. Yamaguchi, Phys. Lett. B{\bf 485}, 388 (2000)[arXiv:hep-
ph/0005214];
F. Boudjema, A. Semenov, D. Temes, Phys. Rev. {\bf D72}, 055024 (2005) [arXiv:hep-ph/0507127];
C. Boehm, J. Orloff, P. Salati, Phys. Lett. {\bf B641}, 247 (2006) [arXiv:astro-ph/0607437];
L. Bergstrom, T. Bringmann, M. Eriksson, M. Gustafsson, JCAP {\bf 0504}, 004 (2005) [arXiv:hep-ph/0412001];
etc..
\bibitem{douglas}
N. Seiberg, and E. Witten, JHEP, {\bf 09}, 032 (1999).
\bibitem{minimallength}
S. Doplicher, K. Fredenhagen, and J. E. Roberts, Phys. Lett. B{\bf 331}, 39 (1994); Commun. Math. Phys. {\bf 172}, 187 (1995).
\bibitem{UVIR}
S. Minwalla, M. van Raamsdonk and N. Seiberg, JHEP {\bf 0002}, 020
(2000);
 A. Matusis, L. Susskind and N. Toumbas, JHEP {\bf 0012}, 002 (2000).
 \bibitem{lorentz}
 M. Chaichian, P. Kulish, K. Nishijima, A. Tureanu, Phys. Lett. B{\bf 604},98 (2004);
M. Chaichian, P. Presnajder, A. Tureanu, Phys. Rev. Lett. {\bf 94}, 151602 (2005).
%\bibitem{charge}
%M.M. Sheikh-Jabbari, Phys. Rev. Lett. {\bf 84}, 5265 (2000).
\bibitem{bound}
I. Mocioiu, M. Pospelov, R. Roiban, Phys. Lett. B{\bf 489}, 390 (2000); M. Chaichian, M. M. Sheikh-Jabbari, A. Tureanu, Phys. Rev. Lett. {\bf 86}, 2716 (2001); J.L. Hewett, F.J. Petriello, T.G. Rizzo, Phys. Rev.
D {\bf 64}, 075012 (2001); S. M. Carrool, J. A. Harvey, V. A. Kostelecky, C. D. Lane, and T.
Okamoto, Phys. Rev. Lett. {\bf 87}, 141601 (2001).
\bibitem{nph}
H. Grosse and Y. Liao, Phys. Lett. B {\bf 520}, 63 (2001) [hep-ph/0104260]; H. Grosse and Y. Liao, Phys. Rev. D {\bf 64}, 115007 (2001) [hep-ph/0105090];
 P. Schupp, J. Trampetic, J. Wess and G. Raffelt, Eur. Phys. J. C {\bf 36}, 405 (2004)
[hep-ph/0212292]; M. Haghighat, M.M. Ettefaghi and M. Zeinali, Phys. Rev. D {\bf 73}, 013007 (2006) [hep-ph/0511042];
M.M. Ettefaghi and M. Haghighat, Phys. Rev. D {\bf 77},  056009 (2008) [arXiv:0712.4034], M. M. Ettefaghi and T. Shakouri, JHEP {\bf 11}, 131 (2010);
 R. Horvat, D. Kekez, P. Schupp, J. Trampetic, and J. You, Phys. Rev. D {\bf 84},  045004 (2011); M. Haghighat,
Phys. Rev. D{\bf 79}, 025011 (2009).
 \bibitem{fermi} A.A. Abdo et al., Phys. Rev. Lett. 104  091302 (2010)
[arXiv:1001.4836].
\bibitem{nogo}
M. Chaichian, P. Presnajder, M. M. Sheikh-Jabbari, and A. Tureanu,
Phys. Lett. B {\bf 526}, 132 (2002).
\bibitem{nNCSM} M. Chaichian, P. Presnajder, M. M. Sheikh-Jabbari, and A.
Tureanu, Eur. Phys. J. C{\bf 29}, 413(2003).
\bibitem{swmap}
J. Madore, S. Schraml, P. Schupp, and J. Wess, Eur. Phys. J. C {\bf
16}, 161 (2000); B. Jur\v co, S. Schraml, P. Schupp, and J. Wess,
Eur. Phys. J. C {\bf 17}, 521 (2000); B. Jur\v co, L. M\"oller, S.
Schraml, P. Schupp, and J. Wess, Eur. Phys. J. C {\bf 21}, 383
(2001); L. M\"oller, J. High Energy Phys. {\bf 10}, 063 (2004);
M.~M.~Ettefaghi and M.~Haghighat, Phys. Rev. D{\bf 75}, 125002
(2007).
\bibitem{sm}
X.~Calmet, B.~Jur\v co, P.~Schupp, J.~Wess and M.~Wohlgenannt, Eur.\
Phys.\ J. {\bf C23}, 363 (2002); B. Meli\'{c}, K.
Passek-Kumeri\v{c}ki, J. Trampeti\'{c}, P. Schupp and M.
Wohlgenannt, Eur.\ Phys.\ J. {\bf C42}, 483(2005).
\bibitem{bypass}
 M. Chaichian, P. Presnajder, M. M. Sheikh-Jabbari, and A. Tureanu, Phys. Lett. B{\bf 683}, 55 (2010).
\end{thebibliography}
\end{document}